\documentclass[epj]{svjour}
\usepackage{graphicx}
\begin{document}

\title{Density-Constrained Time-Dependent Hartree-Fock\\
        Calculation of $^{16}$O+$^{208}$Pb Fusion Cross-Sections}
\author{A.S. Umar\inst{1}\thanks{\emph{e-mail:} umar@compsci.cas.vanderbilt.edu} \and V.E. Oberacker\inst{1}}

\authorrunning{Umar, Oberacker}
\titlerunning{DC-TDHF Calculation of $^{16}$O+$^{208}$Pb Fusion}

\institute{Department of Physics and Astronomy, Vanderbilt University, Nashville, Tennessee 37235, USA}

\date{Received: \today / Revised version: date}

\abstract{
We present a fully microscopic study of the $^{16}$O+$^{208}$Pb fusion
using the density-constrained time-dependent Hartree-Fock theory.
The calculated fusion cross-sections are in good agreement with the
experimental data for the entire energy range indicating that the
incorporation of dynamical effects is crucial in describing heavy-ion
fusion.
} 

\maketitle

\section{Introduction}

Heavy-ion fusion reactions are a sensitive probe of the size, shape, and structure
of atomic nuclei as well as the collision dynamics. Fusion studies using neutron-rich
nuclei are becoming increasingly available. Examples include the
experiments with heavy neutron-rich $^{132}$Sn beams on $^{64}$Ni~\cite{Li03,Li07},
and the fusion measurements for the $^{64}$Ni+$^{64}$Ni system~\cite{Ji04}.
What characterized these experiments was the measurement of the fusion cross-sections
at very low bombarding energies. Recently, fusion cross-sections for the $^{16}$O+$^{208}$Pb
system were also measured at extreme sub-barrier energies~\cite{Das07}.
Another experimental frontier for fusion is the synthesis of superheavy nuclei in cold and hot fusion
reactions~\cite{Ho02,Og04,Lo07,II05}.

To date quantitative theoretical studies of fusion cross-sections have largely been limited
to phenomenological methods, mainly the coupled-channels approach. Specifically, various
effects such as neutron transfer~\cite{Li07}, compression potential~\cite{Esb06}, modifications of the inner turning
points via the approximate inclusion of neck formation~\cite{IH07a,IH07b}, and complex potentials~\cite{Esb08} were introduced,
in addition to standard channel couplings, to address the recent data at extreme sub-barrier energies.
This is surprising since the $^{16}$O+$^{208}$Pb system should be an ideal system for
coupled-channel calculations. A number of papers have been devoted exclusively to the
$^{16}$O+$^{208}$Pb system~\cite{Esb08,HW07}.
This underscores the fact that
while phenomenological methods provide a useful
and productive means for quantifying multitudinous fusion data it is desirable to
make contact with the microscopic theories of nuclear structure and reactions.

In the absence of a true many-body theory for calculating sub-barrier fusion all practical
calculations are reduced to the calculation of a potential barrier between the interacting
nuclei and a subsequent calculation of tunneling through the barrier.
Many of the phenomenological methods use frozen densities for each nuclei and ignore all
of the dynamical effects as well as the formation of a neck during the nuclear overlap.
It has been demonstrated that for deep sub-barrier energies the inner part of the
potential barrier plays a very important role~\cite{IH07a,IH07b}. While the outer part
of the barrier is largely determined by the early entrance channel properties of the
collision, the inner part of the potential barrier is strongly sensitive to dynamical
effects such as particle transfer and neck formation.
Recently, an inversion method has been devised
\cite{HW07} to obtain inter-nucleus potentials directly from the
fusion data for the $^{16}$O+$^{208}$Pb system. In their conclusion the authors propose
that dynamical effects, such as neck formation and coordinate
dependent mass may be responsible for the modification of the
ordinary potentials used in the coupled-channel calculations.

It is generally acknowledged that the time-dependent Hartree-Fock (TDHF) theory provides a
useful foundation for a fully microscopic many-body theory of low-energy heavy-ion collisions~\cite{Ne82}.
The success of the TDHF method is predicated on the expectation that the Pauli principle plays an important role in
simultaneously building up a time-dependent mean-field and suppressing the propagation of the
strong $N-N$ interaction terms. With the advent of computer technology it
has now become feasible to perform TDHF calculations with no symmetry assumptions regarding the
collision process and the effective interaction, and with much more accurate numerical methods~\cite{UO06}.
At the same time the quality of effective interactions has also been substantially improved~\cite{CB98}.
In this sense, a description of the heavy-ion collisions via the TDHF method may provide a way to incorporate
the dynamical effects such as mass transfer and neck formation into the calculation of the
ion-ion potential.

In Section~2 we outline the main features of one such method, the density-constrained
TDHF (DC-TDHF) method~\cite{UO06b}.
The DC-TDHF method was first used to study the fusion reactions of the
$^{64}$Ni+$^{132}$Sn system~\cite{UO06d,UO07a}. A re-measurement of the lowest energy data point~\cite{Li08}
has moved this point exactly onto the DC-TDHF result. We have also performed calculations for the
$^{64}$Ni+$^{64}$Ni system~\cite{UO08a} and showed that an excellent agreement with data is
possible. As the DC-TDHF method contains no parameters or normalization its primary success depends
on the quality of the description of the ground state properties within the Hartree-Fock framework
and the description of the early stages of the ion-ion reaction by the TDHF theory. In this sense
we expect the $^{16}$O+$^{208}$Pb system to be a good test since most Skyrme parametrizations include
these nuclei in the fitting procedure.
In Section~2 we also discuss
the calculation of ion-ion separation distance, coordinate-dependent mass, and calculation
of fusion cross-sections. In Section~3 the calculations for the $^{16}$O+$^{208}$Pb system
are presented. Final conclusions are given in Section~4.

\section{Formalism}
\subsection{DC-TDHF method}
Recently, we have developed a new method to extract ion-ion interaction potentials directly from
the TDHF time-evolution of the nuclear system.
In the DC-TDHF approach~\cite{UO06b}
the TDHF time-evolution takes place with no restrictions.
At certain times during the evolution the instantaneous density is used to
perform a static Hartree-Fock minimization while holding the neutron and proton densities constrained
to be the corresponding instantaneous TDHF densities. In essence, this provides us with the
TDHF dynamical path in relation to the multi-dimensional static energy surface
of the combined nuclear system. The advantages of this method in comparison to other mean-field
based microscopic methods such as the constrained Hartree-Fock (CHF) method are obvious. First,
there is no need to introduce artificial constraining operators which assume that the collective
motion is confined to the constrained phase space, second the static adiabatic approximation is
replaced by the dynamical analogue where the most energetically favorable state is obtained
by including sudden rearrangements and the dynamical system does not have to move along the
valley of the potential energy surface. In short we have a self-organizing system which selects
its evolutionary path by itself following the microscopic dynamics.
Some of the effects naturally included in the DC-TDHF calculations are: neck formation, mass exchange,
internal excitations, deformation effects to all order, as well as the effect of nuclear alignment
for deformed systems.
In the DC-TDHF method the ion-ion interaction potential is given by
\begin{equation}
V(R)=E_{\mathrm{DC}}(R)-E_{\mathrm{A_{1}}}-E_{\mathrm{A_{2}}}\;,
\label{eq:vr}
\end{equation}
where $E_{\mathrm{DC}}$ is the density-constrained energy at the instantaneous
separation $R(t)$, while $E_{\mathrm{A_{1}}}$ and $E_{\mathrm{A_{2}}}$ are the binding energies of
the two nuclei obtained with the same effective interaction.
In writing Eq.~(\ref{eq:vr}) we have introduced the concept of an adiabatic reference state for
a given TDHF state. The difference between these two energies represents the internal energy. 
The adiabatic reference state is the one obtained via the density constraint calculation, which 
is the Slater determinant with lowest energy for the given density with vanishing current
and approximates the collective potential energy~\cite{CR85}.
We would like to
emphasize again that this procedure does not affect the TDHF time-evolution and
contains no {\it free parameters} or {\it normalization}.

In addition to the ion-ion potential it is also possible to obtain coordinate
dependent mass parameters. One can compute the ``effective mass'' $M(R)$
using the conservation of energy
\begin{equation}
M(R)=\frac{2[E_{\mathrm{c.m.}}-V(R)]}{\dot{R}^{2}}\;,
\label{eq:mr}
\end{equation}
where the collective velocity $\dot{R}$ is directly obtained from the TDHF evolution and the potential
$V(R)$ from the density constraint calculations.

\subsection{Calculation of $R$}
In practice, TDHF runs are initialized with energies above the Coulomb barrier at some
large but finite separation. The two ions are boosted with velocities obtained by assuming
that the two nuclei arrive at this initial separation on a Coulomb trajectory.
Initially the nuclei are placed such that the point $x=0$ in the $x-z$ plane is the center of mass.
During the TDHF dynamics the ion-ion separation distance is obtained by constructing a dividing plane between the
two centers and calculating the center of the densities on the left and right halves of this
dividing plane. The coordinate $R$ is the difference between the two centers.
The dividing plane is determined by finding the point at which the tails of the two
densities intersect each other along the $x$-axis. Since the actual mesh used in the
TDHF calculations is relatively coarse we use a cubic-spline interpolation to interpolate
the profile in the x-direction and search for a more precise intersection value.
This procedure has been recently
described in Ref.~\cite{DD-TDHF} in great detail. We have confirmed with the authors
of Ref.~\cite{DD-TDHF} that our $R$ values are in exact agreement with their calculations for
the $^{16}$O+$^{208}$Pb system.

The above procedure works well after the formation of the neck between the two ions. However,
at some point when the two ions are substantially overlapping the procedure starts to fail.
While this does not occur for the $R$ values used in this calculation, it may be desirable to
find an alternate way of determining the $R$ value. We have found that if one defines the
ion-ion separation as $R=R_0\sqrt{|Q_{20}|}$, where $Q_{20}$ is the mass quadrupole moment
for the entire system, calculated by using the collision axis as the symmetry axis,
and $R_0$ is a scale factor determined to give the correct initial
separation distance at the start of the calculations. Calculating $R$ this way yields
almost identical results to the previous procedure until that procedure begins to fail
and continues smoothly after that point. We have compared the two methods for the
$^{16}$O+$^{16}$O system for which the dividing plane is essentially at the origin
for all $R$ values. We find that the quadrupole method gives the correct $R$ value even after
substantial overlap. Of course the minimum value of $R$ calculated this way is never zero
but is determined by the quadrupole moment of the composite system. In this sense the
above approach may be useful for fission studies.

\subsection{Fusion cross-section}
We now outline the calculation of the total fusion cross section using an
arbitrary coordinate-dependent mass $M(R)$. Starting from the classical
Lagrange function 
\begin{equation}
L(R,\dot{R})=\frac{1}{2} M(R) {\dot{R}^{2}} - V(R) \;,
\label{eq:Lag}
\end{equation}
we obtain the corresponding Hamilton function
\begin{equation}
H(R,P)=\frac{P^2}{2 M(R)} + V(R) \;,
\label{eq:Ham1}
\end{equation}
where the canonical momentum is given by $P=M(R){\dot{R}}$.
Following the standard quantization procedure for the kinetic energy 
in curvilinear coordinates~\cite{Sp59}
\begin{equation}
T = \frac{-\hbar^2}{2} \left[ g^{-{\frac{1}{2}}} \frac{\partial}{\partial q^{\mu}} 
    g^{{\frac{1}{2}}} g^{\mu \nu} \frac{\partial}{\partial q^{\nu}}\right] \;,
\label{eq:Tgen}
\end{equation}
where $g_{\mu \nu}(q)$ denotes the metric tensor and $g^{\mu \nu}(q)$ the reciprocal tensor,
one obtains the quantized Hamiltonian
\begin{equation}
H(R,\hat{P})=\frac{1}{2} \left[ M(R)^{-{\frac{1}{2}}} \hat{P} M(R)^{-{\frac{1}{2}}} \hat{P} \right] + V(R) \;.
\label{eq:Ham2}
\end{equation}
with the momentum operator $\hat{P} = -i \hbar d/dR$.
The total fusion cross cross-section
\begin{equation}
\sigma_f = \frac{\pi}{k^2} \sum_{L=0}^{\infty} (2L+1) T_L\;,
\label{eq:sigfus}
\end{equation}
can be obtained by calculating the potential barrier penetrabilities $T_L$
from the Schr\"odinger equation for the relative motion coordinate $R$
using the Hamiltonian~(\ref{eq:Ham2}) with an additional centrifugal potential
\begin{equation}
\left [ H(R,\hat{P}) + \frac{\hbar^2 L(L+1)}{2 M(R) R^2}
 - E_\mathrm{c.m.} \right] \psi_L(R) = 0 \;.
\label{eq:Schroed1}
\end{equation}

Alternatively, instead of solving the Schr\"odinger equation with coordinate dependent
mass parameter $M(R)$ for the heavy-ion potential $V(R)$, we can instead use the constant
reduced mass $\mu$ and transfer the coordinate-dependence of the mass to a scaled
potential $U(\bar{R})$ using the well known coordinate scale transformation~\cite{GRR83}
\begin{equation}
d\bar{R}=\left(\frac{M(R)}{\mu}\right)^{\frac{1}{2}}dR\;.
\label{eq:mrbar}
\end{equation}
Integration of Eq.~(\ref{eq:mrbar}) yields
\begin{equation}
\bar{R}= f(R) \ \ \ \Longleftrightarrow \ \ \ R=f^{-1}(\bar{R})\;.
\label{eq:rbar}
\end{equation}
As a result of this point transformation, both the classical
Hamilton function, Eq.~(\ref{eq:Ham1}), and the corresponding quantum mechanical
Hamiltonian, Eq.~(\ref{eq:Ham2}), now assume the form
\begin{equation}
H(\bar{R},\bar{P})=\frac{\bar{P}^2}{2 \mu} + U(\bar{R}) \;,
\label{eq:Ham3}
\end{equation}
and the scaled heavy-ion potential is given by the expression
\begin{equation}
U(\bar{R}) = V(R) = V(f^{-1}(\bar{R})) \;.
\label{eq:Urbar}
\end{equation}

\section{Results}
In our numerical calculations for the $^{16}$O+$^{208}$Pb system we have chosen a
Cartesian box which is $60$~fm along the collision axis and $30$~fm in
the other two directions. The two nuclei are placed at an initial separation of $24$~fm.
Calculations are done in 3-D geometry and using the full Skyrme force (SLy4)~\cite{CB98}
as described in
Ref.~\cite{UO06}. The numerical accuracy of the static binding energies and the deviation
from the point Coulomb energy in the initial state of the collision dynamics is on the order
of $50-150$~keV. We have performed density constraint calculations at every $10$~fm/c interval or
a total of $40$ calculations. The accuracy of the density constraint calculations are
commensurate with the accuracy of the static calculations.
\begin{figure}[!htb]
\includegraphics*[scale=0.42]{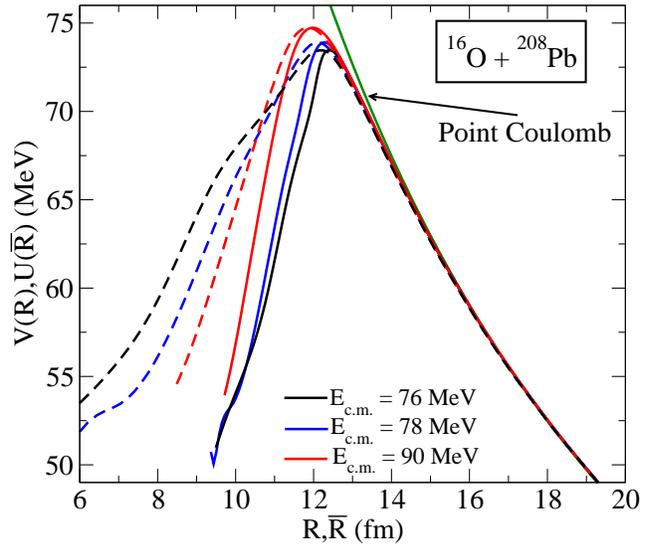}
\caption{\label{fig:vrb} (Colour on-line) Potential barriers, $V(R)$, obtained from density constrained
TDHF calculations using Eq.~\protect(\ref{eq:vr}) at three different energies,  $E_\mathrm{c.m.}=76$~MeV
(black solid curve), $E_\mathrm{c.m.}=78$~MeV
(blue solid curve), and $E_\mathrm{c.m.}=90$~MeV (red solid curve). The three dashed curves
correspond to the transformed potential in Eq.~\protect(\ref{eq:Urbar}) using the coordinate dependent
masses. Also shown is the point Coulomb potential.}
\end{figure}

\begin{figure}[!htb]
\includegraphics*[scale=0.42]{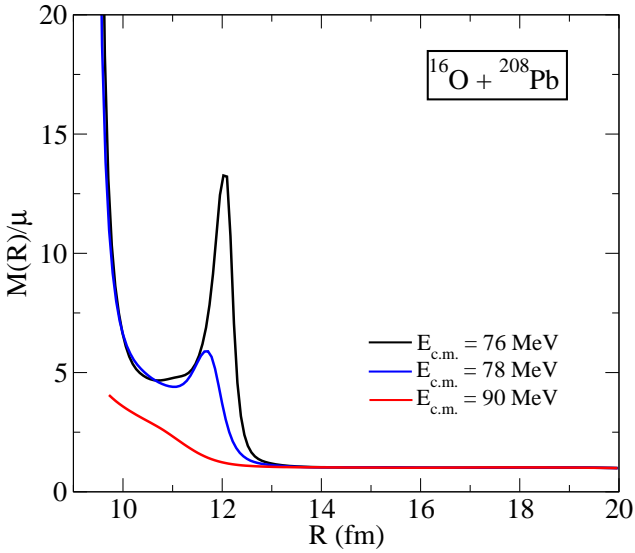}
\caption{\label{fig:mr} (Colour on-line) Coordinate dependent mass $M(R)$ scaled by the
constant reduced mass $\mu$, obtained from
Eq.~\protect(\ref{eq:mr}), at three different energies,  $E_\mathrm{c.m.}=76$~MeV
(black solid curve), $E_\mathrm{c.m.}=78$~MeV
(blue solid curve), and $E_\mathrm{c.m.}=90$~MeV (red solid curve).}
\end{figure}

In Fig.~\ref{fig:vrb} we show the ion-ion potentials obtained using Eq.~(\ref{eq:vr})
calculated at three different energies, $E_\mathrm{c.m.}=76$~MeV (black solid curve),
$E_\mathrm{c.m.}=78$~MeV (blue solid curve), and 
$E_\mathrm{c.m.}=90$~MeV (red solid curve). The same energy dependence was also observed
in the DD-TDHF calculations of Ref.~\cite{DD-TDHF} and seems to be more prevalent for
heavier systems. For example in $^{16}$O+$^{16}$O system we saw a very small
energy dependence even though the energy was tripled from $12$~MeV to $36$~MeV.
As can be expected, at lower energies we observe more dynamical effects such as
mass transfer, whereas the high energy results seem to approach the frozen-density
limit~\cite{DD-TDHF}. This is also apparent in the coordinate dependent mass of
Eq.~(\ref{eq:mr}) shown in Fig.~\ref{fig:mr} for three different energies, $E_\mathrm{c.m.}=76$~MeV
(black solid curve), $E_\mathrm{c.m.}=78$~MeV
(blue solid curve), and $E_\mathrm{c.m.}=90$~MeV (red solid curve).
The $R$-dependence of this mass at lower energies is
very similar to the one found in CHF calculations~\cite{GRR83}.
On the other hand, at higher energies the coordinate dependent mass essentially becomes flat,
which is again a sign that most dynamical effects are contained at
lower energies.
The peak at small $R$ values is
due to the fact that the center-of-mass energy is above the barrier and the
denominator of Eq.~(\ref{eq:mr}) becomes small due to the slowdown of the ions.
We have used the coordinate dependent masses shown in Fig.~\ref{fig:mr} to obtain the
scaled potentials $U(\bar{R})$ of Eq.~(\ref{eq:Urbar}). These potentials are shown
as the dashed curves in Fig.~\ref{fig:vrb}. As we see the coordinate dependent mass
only changes the inner parts of the barriers for all energies. Furthermore, the
effect is largest for the lowest energy collision and diminishes as we increase
the collision energy.

\begin{figure}[!htb]
\includegraphics*[scale=0.40]{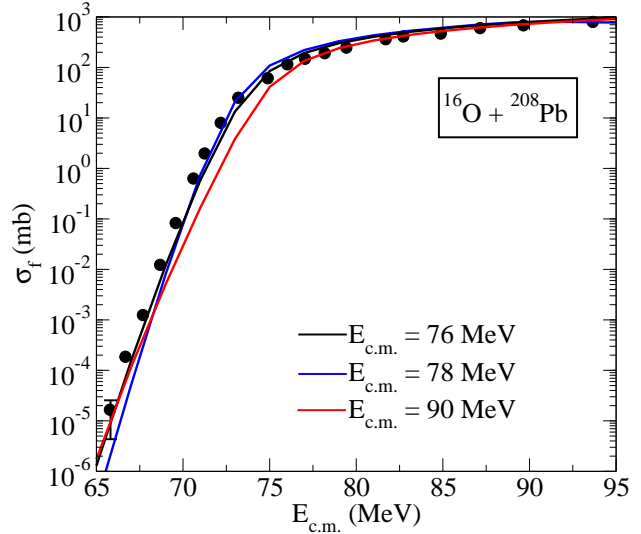}
\caption{\label{fig:fus} (Colour on-line) Total fusion cross section as a function of $E_{\mathrm{c.m.}}$.
Three separate theoretical cross section calculations are shown, based on the energy-dependent
DC-TDHF heavy-ion potentials $V(R)$ and coordinate dependent masses $M(R)$ at energies
$E_{\mathrm{c.m.}}=76,78,90$ MeV, from which the scaled potentials $U(\bar{R})$ were obtained.
The experimental data (filled circles) are taken from Ref.~\protect\cite{Das07}.}
\end{figure}

We have obtained the fusion cross sections by numerical integration of the
Schr\"odinger equation using the constant reduced mass $\mu$ and the scaled potential
$U(\bar{R})$.
The fusion barrier penetrabilities $T_L(E_{\mathrm{c.m.}})$
are obtained by numerical integration of the two-body Schr\"odinger equation
using the {\it incoming wave boundary condition} (IWBC) method.
IWBC assumes that once the minimum of the potential is reached fusion will
occur. In practice, the Schr\"odinger equation is integrated from the potential
minimum, $R_\mathrm{min}$, where only an incoming wave is assumed, to a large asymptotic distance,
where it is matched to incoming and outgoing Coulomb wavefunctions. The barrier
penetration factor, $T_L(E_{\mathrm{c.m.}})$ is the ratio of the
incoming flux at $R_\mathrm{min}$ to the incoming Coulomb flux at large distance.
Here, we implement the IWBC method exactly as it is
formulated for the coupled-channel code CCFULL described in Ref.~\cite{HR99}.
This gives us a consistent way for calculating cross-sections at above and below
the barrier energies. The resulting cross-sections are shown in Fig.~\ref{fig:fus}.

We observe that all of the scaled barriers give a very good description of the
fusion cross-section at higher energies suggesting these cross-sections are
primarily determined by the barrier properties in the vicinity of the barrier
peak. Naturally, the extreme sub-barrier cross-sections are influenced by what
happens in the inner part of the barrier and here the dynamics and
consequently the coordinate dependent mass
becomes very important. As we observe in Fig.~\ref{fig:fus} we get
a very good agreement with experiment for the barriers obtained at the lowest
two collision energies, whereas the $E_\mathrm{c.m.}=90$~MeV curve underestimates
the cross-section at lower energies. Although not shown in Fig.~\ref{fig:fus} the
cross-sections obtained using the the unscaled potentials $V(R)$ and a constant
reduced mass $\mu$ also agree well with the data for $E_\mathrm{c.m.}>77$~MeV
but either significantly over-estimate or under-estimate the cross-section at
lower energies.

\section{Conclusions}
In summary, we have used the fully microscopic DC-TDHF method to study $^{16}$O+$^{208}$Pb fusion.
The standard, parameter free approach of DC-TDHF yields potential barriers that can
accurately reproduce the fusion cross-sections. The DC-TDHF approach has now been applied
to study a number of systems with very promising results. Our results also agree with
the conclusion of Ref.~\cite{HW07} that for the proper description of fusion cross-sections
at deep sub-barrier energies the dynamical effects such as neck formation and mass transfer
must be included to modify the inner part of the potential barrier. Also, interesting
are the implications of the energy dependent barriers as was also discussed extensively
in Ref.~\cite{DD-TDHF}.
These findings underscore the fact that
additional modifications needed for phenomenological methods to explain the fusion
cross-sections of the $^{16}$O+$^{208}$Pb system may largely be due to the inadequacy
of the approximations made in treating the nuclear dynamics.

This work has been supported by the U.S. Department of Energy under grant No.
DE-FG02-96ER40963 with Vanderbilt University.

\end{document}